\documentclass[conference]{IEEEtran}
\usepackage{url}
\usepackage{footnote}
\ifCLASSINFOpdf
\else
\fi

\usepackage{amssymb}
\usepackage{graphicx}
\usepackage{amssymb}
\usepackage{amsmath}
\usepackage{color,rotating}
\newtheorem{Theorem}{Theorem}

\newcommand{\cqfd}{\mbox{}\nolinebreak\hfill\rule{2mm}{2mm}\medbreak\par}

\begin{document}
%
\title{Optimizing Scanning Strategies: Selecting Scanning Bandwidth in Adversarial RF Environments}

\author{\IEEEauthorblockN{Andrey Garnaev}
\IEEEauthorblockA{
St. Petersburg State University, Russia\\
Email: garnaev@yahoo.com} \and \IEEEauthorblockN{Wade Trappe,
Chun-Ta Kung}
\IEEEauthorblockA{WINLAB and the Electrical and Computer Engineering Department\\
Rutgers University,
USA\\
Email: trappe@winlab.rutgers.edu, chuntakung@gmail.com}}


%


\maketitle

\begin{abstract}
In this paper we investigate the problem of designing a spectrum scanning strategy to detect an
 intelligent Invader who wants to utilize spectrum
undetected for his/her unapproved purposes.  To deal with this
problem we apply game-theoretical tools. We model the situation as
a game between a Scanner and an Invader where the Invader faces a
dilemma: the more bandwidth the Invader attempts to use leads to a
larger payoff if he is not detected, but at the same time also
increases the probability of being detected and thus fined.
Similarly, the Scanner faces a dilemma: the wider the bandwidth
scanned, the higher the probability of detecting the Invader, but
at the expense of increasing the cost of building the scanning
system.  The equilibrium strategies are found explicitly and
reveal interesting properties. In particular, we have found a
discontinuous dependence of the equilibrium strategies on the
network parameters, fine and the type of the Invader's award. This
discontinuity on fine means that the network provider has to take
into account a human factor since some threshold values of fine
could be very sensible for  the Invader, while in other situations
simply increasing the fine has minimal deterrence impact. Also we
show how different reward types for the Invader (e.g. motivated by
using different type of application, say, video-streaming or
downloading files) can be incorporated into scanning strategy to
increase its efficiency

\footnote[1]{This is  the last draft version of the paper. Revised version of the paper accepted by CrownCom 2013 can be found at
\url{http://ieeexplore.ieee.org/xpl/login.jsp?tp=&arnumber=6636809&url=http\%3A\%2F\%2Fieeexplore.ieee.org\%2Fxpls\%2Fabs_all.jsp\%3Farnumber\%3D6636809}
}
.

\end{abstract}

\section{Introduction}
Over the last few decades, the increasing demand for wireless
communications has motivated the exploration for more efficient
usage of spectral resources (\cite{1391031,819467}). In
particular, it has been noticed that there are large portions of
spectrum that are severely under-utilized \cite{akyildiz2006next}.
Recently, cognitive radio technologies (CR) have been proposed as
a means to intelligently use such spectrum opportunities by
sensing the radio environment and exploiting available spectrum
holes for secondary usage\cite{fette2009cognitive}. In CR systems,
secondary users are allowed to ``borrow (or lease)'' the usage of
spectrum from primary users (licensed users), as long as they do
not hinder in the proper operation of the primary users'
communications. Unfortunately, as we move to make CR technologies
commercial, which will allow secondary users to access spectrum
owned by primary users, we will face the inevitable risk that
adversaries will be tempted to use CR technology for illicit and
selfish purposes \cite{liu2009aldo}. If we imagine an unauthorized
user (Invader) attempting to sneak usage of spectrum without
obeying proper regulations or leasing the usage of the spectrum,
the result will be that both legitimate secondary users and
primary users will face unexpected interference, resulting in
significant performance degradation across the system.

The challenge of enforcing the proper usage of spectrum requires
the notion of a ``spectrum policing agent'', whose primary job is
to ensure the proper usage of spectrum and identify anomalous
activities occurring within the spectrum\cite{liu2009aldo}.  As a
starting point to being able to police the usage of spectrum, we
must have the ability to scan spectrum and effectively identify
anomalous activities. Towards this objective, there has been
several research efforts in signal processing techniques that can
be applied to the spectrum scanning problem. For example, in
\cite{verdu1998multiuser,van2004detection}, the author's presented
methods for detecting a desired signal contained within
interference. Similarly, detection of unknown signals in noise
without prior knowledge of authorized users was studied in
\cite{digham2007energy,urkowitz1967energy}. As another example, in
\cite{liu2009aldo}, the authors proposed a method to detect
anomalous transmission by making use of radio propagation characteristics.
In \cite{GTK2012} authors investigated what
impact on spectrum scanning can have information about  the
over-arching application that a spectrum thief might try to run.

However, these works tend to not examine the important ``interplay'' between
the two participants inherent in the problem-- the Invader, who is smart
and will attempt to use the spectrum in a manner to minimize the chance
of being detected and fined, while also striving to maximize
the benefit he/she receives from illicit usage of this spectrum;
and the Scanner, who must be smart and employ a strategy that
strategically maximizes the chance of detecting and fining the smart Invader,
with minimal cost. This challenge is made more difficult by the complexity of
the underlying scanning problem itself: there will be large swaths of bandwidth to scan,
and the system costs (e.g. analog-to-digital conversion, and the computation associated with running signal
classifiers) associated with scanning very wide bandwidth makes
it impossible to scan the full range of spectrum in a single instance.
Consequently, it is important to understand the strategic dynamics that exist between the Scanner and the Invader,
while also taking into account the underlying costs and benefits that exist for each participant.
This paper examines the interactions between the Scanner and Invader by
formulating the problem using game theory. We find the optimal scanning strategy by selecting the scanning
(and, similarly, the invading) bandwidth that should be employed in spectrum scanning.

The organization of this paper is as follows: in Section
\ref{sec:Problem}, we first define the problem by formulating a
two-step  game in terms of payoff and cost functions: in the first step, each player chooses their bandwidths; while in the second step each player uses the first step results to arrive at equilibrium strategies. We supply saddle point strategies for the game's first step. To gain insight into the problem, for the second step, in Section \ref{sec:Linear}, we outline a linearized model of detection probability and arrive at the corresponding best strategies for each player in Section \ref{sec:Best
Response}.  We then explicitly obtain the equilibrium
strategies  in Sections \ref{sec:Nash Equilibrium}  and
\ref{sec:Nash Equilibrium_Unknown} for cases involving complete and incomplete
knowledge of the Invader's reward type. In Section \ref{sec:Numerical illustrations}
numerical illustrations are supplied. Finally, in Sections
\ref{sec:Discussion} and \ref{sec:Appendix} discussions,
conclusions and the proofs of the announced results are offered to
close the paper.

\section{\label{sec:Problem}
Formulation of the Scanning Problem as a Two-Step Game}

In this section we set up our problem formulation. Our formulation
of the spectrum scanning problem involves  two players: the
Scanner and the Invader. The Scanner, who is always present in the
system, scans a part of the band of frequencies that are to be
monitored, in order to prevent illegal usage by a potential
Invader of the primary (Scanner) network's ownership of this band.
We assume that the amount of bandwidth that needs to be scanned is
much larger than is possible using a single scan by the Scanner,
and hence the Scanner faces a dilemma: the more bandwidth that is
scanned, the higher the probability of detecting the Invader, but
at the expense of increasing the cost of the RF scanning system.

We assume that if the Scanner scans a particular frequency band
$I_S$ and the Invader uses the band $I_I$ then the invasion will
be detected with certainty if $I_S\cap I_I\not=\emptyset$, and it
will not be detected otherwise. Without loss of generality we can
assume that the size of the protected frequency band is normalized
to 1. The Invader wants to use spectrum undetected  for some
illicit purpose. We consider two scenarios: (a) The reward for the
Invader is related to the width of the frequency band he uses if
he is undetected. If he is detected he will be fined. So, the
Invader faces a dilemma: the more bandwidth he tries to use yields
a larger payoff if he is not detected but also it increases the
probability of being detected and thus to be fined, (b) The award
for the Invader is unknown to the Scanner: he only knows whether it
is related to the width of the frequency band the
Invader uses, or not.

We formulate this problem as a two step game in the following two
subsections.

\subsection{Formulation of the Problem in the First Step of the
Game}

In the first step of the game the Scanner selects the band
$B_S=[t_S,t_S+x]\subseteq [0,1]$ with a fixed upper bound of
frequency width $x$ to scan i.e. $t_S\leq 1-x$. The Invader
selects the band $B_I=[t_I,t_I+y] \subseteq [0,1]$  with a fixed
upper bound frequency width $y$ to intrude, i.e. $t_I\leq 1-y$.
So, $B_S$ and $B_I$ are pure strategies for the Scanner and the
Invader. The Scanner's payoff $v(B_S,B_I)$ is 1 if the Invader is
detected (i.e. $[t_S,t_S+x]\cap [t_I,t_I+y]\not=\emptyset$) and
his payoff is zero otherwise. The goal of the Scanner is to
maximize his payoff meanwhile the Invader wants to minimize it.
So, the Scanner and the Invader play a zero-sum game. The saddle
point (equilibrium) of the game is a couple of strategies
$(B_{S*},B_{I*})$ such that for each strategies $(B_{S},B_{I})$
the following inequalities hold \cite{O1982}:
$$
v(B_{S},B_{I*})\leq v:=v(B_{S*},B_{I*})\leq v(B_{S*},B_{I}),
$$
where $v$ is the value of the game. It is clear that the game does
not have a saddle point in the pure strategy if $x+y\leq 1$. To
find the saddle point we have to extend the game by mixed
strategies, where we assign a probability distribution over pure
strategies. Then instead of the payoff $v$ we have its expected
value. The game has a saddle point in mixed strategies, and let
$P(x,y)$ be the value of the game. Then $P(x,y)$ is the maximal
detection probability of the Invader under worst conditions.

\subsection{Formulation of the Problem in the Second Step of the
Game}

In the second  step of the game the rivals knowing their
equilibrium strategies from the first step as well as  detection
probability $P(x,y)$, want  to find the equilibrium frequency widths $x$ and $y$.
We here consider two sub-scenarios: (a) the
Invader's type is known: namely, it is known how the reward for the
Invader is related to the width of the frequency band he uses if
he is undetected, (b) the Invader's type is unknown: instead, there is only a chance that the Invader reward is related to the width in use, else it is not related. Different type of rewards can be
motivated by using different type of application (say,
file-download or streaming video \cite{GTK2012}).

\subsubsection{Invader reward is related to the bandwidth used}
 A  strategy for the Scanner is to scan a width of
frequency of size $x\in [a,b]$, and a strategy for the Invader is
to employ a width of frequency of size $y\in [a,c]$, where
$c<b<1/2$. So, we assume that the Invader's technical
characteristics is not better than the Scanner's ones.

If the Scanner and the Invader use the strategies $x$ and $y$,
then the payoff to the Invader is the expected award (which is a
function $U(y)$ of bandwidth $y$ illegally used by the Invader)
minus intrusion expenses (which is a function $C_I(y)$ of
bandwidth $y$) and expected fine $F$ to pay, i.e.
\begin{equation}
\label{eE_a_0} v_I(x,y)= (1-P(x,y))U(y)-FP(x,y) -C_I(y).
\end{equation}

The Scanner wants to detect intrusion  taking into account
scanning expenses and damaged caused by the illegal use of the
bandwidth. For detection he is rewarded by fined $F$ imposed on
the Invader. Thus, the payoff to the Scanner is difference between
the expected reward for detection, and damaged from intrusion into
the bandwidth (which is a function $V(y)$ of bandwidth $y$
illegally used by the Invader) with  the scanning expenses (which
is a function $C_S(x)$ of scanned bandwidth $x$),
\begin{equation}
\label{eE_b} v_S(x,y)=FP(x,y)-V(y)(1-P(x,y))-C_S(x).
\end{equation}
Note that introducing transmission cost is common for CDMA
 \cite{ZSPB2011} and ALOHA networks (\cite{SE2009,GHAA2012}).
We assume that the Scanner and the Invader know fine $F$, cost
functions $C_I$ and $C_S$,  the Scanner and Invader's utilities
$V$ and $U$ as well as low and upper bandwidth bounds $a$, $b$
and $c$. We look for a Nash equilibrium, i.e. for a couple of
strategies $(x_*,y_*)$ such that for any couple of strategies
$(x,y)$ the following inequalities hold \cite{O1982}:
\begin{equation}
\label{e_NE} v_S(x,y_*)\leq v_S(x_*,y_*), \quad v_I(x_*,y)\leq
v_I(x_*,y_*).
\end{equation}

\subsubsection{Unknown Whether the Invader's Award  is Related to the Width of
Band in Use}

In this section we assume that the Invader can be of two types: (a)
with probability $q$ he can be the same as in the previous
section, and so $y$ is his strategy and payoffs are given by
(\ref{eE_a_0}), (b) with probability $1-q$ for him it is just
important to work in the network without being detected. Then, of
course, he will employ the minimal bandwidth allowed, so his
strategy is $y=a$. The payoff to the Scanner is the expected
payoff taking into account the \emph{type} of Invader:
\begin{equation}
\label{eE_b_0} v^E_S(x,y)=qv_S(x,y)+(1-q)v_S(x,a)
\end{equation}
with $v_S(x,y)$ given by (\ref{eE_b}).  Here we also look  for
Nash equilibrium. We assume that the Scanner and the Invader know
(as in the case with complete information) the parameters $F$,
$C_I$, $C_S$, $V$, $U$, $a$, $b$, $c$ as well as the probability
$q$.

\section{\label{sec:Saddle point} Equilibrium  strategies for the first step}
In the following theorem we gives the equilibrium  strategies for
the first step, so for fixed upper bound  width of the rivals.

\begin{Theorem}
\label{th_1} In the first step with fixed  width to scan $x$ and
to invade $y$, the rivals employ uniform tiling behavior. Namely,

(a) Let $1-(x+y)M\leq y$ with
\begin{equation}
\label{e_N} M=\left\lfloor 1/(x+y)\right\rfloor
\end{equation}
where $\lfloor \xi \rfloor$ is the greatest integer less or equal
to $\xi$. Then the Scanner and the Invader will, with equal
probability $1/M$, employ a band of the set $A_{-S}$ and $A_{-I}$
correspondingly.

(b) Let $1-(x+y)M> y$. Then the Scanner and the Invader will, with
equal probability $1/(M+1)$, employ a band of the set $A_{+S}$ and
$A_{+I}$ correspondingly, where

\small
\begin{equation*}
\begin{split}
A_{-S}&=\{[k(x+y)-x,k(x+y)], k=1,...,M\},\\
A_{-I}&=\{[k(x+y)-y-\epsilon (M+1-k),k(x+y)-\epsilon (M-k)],\\
&k=1,..., M\}, \quad 0<\epsilon<x/M,
\\
A_{+S}&=A_{-S}\cup [1-x,1],
\\A_{+I}&=\{[(k-1)(x+y+\epsilon), (k-1)(x+y+\epsilon)+y],\\
&k=1,..., M\}\cup [1-y,1], \quad
0<\epsilon<\frac{1-y-M(x+y)}{M-1}.
\end{split}
\end{equation*}
\normalsize

\noindent The value of the game (detection probability) $P(x,y)$
is given as follows:
\begin{equation}
\label{eP_0} P(x,y)=
\begin{cases}
1/M,& 1-(x+y)M\leq y,\\
 1/(M+1),& 1-(x+y)M > y.\\
\end{cases}
\end{equation}

\end{Theorem}

\section{\label{sec:Second Step}The equilibrium strategy for the second step}
In this section, which is split into four subsections, we find the
equilibrium strategy for the second step explicitly. First in
Subsection~\ref{sec:Linear} we linearize our model to get an explicit
solution, then in Subsection~\ref{sec:Best Response} the best
response strategies are given for known Invader type, and they
are employed in Subsections~\ref{sec:Nash Equilibrium} and
~\ref{sec:Nash Equilibrium_Unknown} to construct equilibrium
strategies for known and unknown Invader types correspondingly.

\subsection{\label{sec:Linear} Linearized  model for the second
step}

In order to get an insight into the problem, we consider a
situation where the detection's probability $P(x,y)$ for $x,y\in
[a,b]$  is approximated by a linear function as follows:
\begin{equation}
\label{eP}
\begin{split}P(x,y)=x+y.
\end{split}
\end{equation}
We assume that the scanning and intrusion cost as well as the
Invader's and Scanner's utilities are linear in the bandwidth
involved, i.e. $C_S(x)=C_Sx$, $C_I(y)=C_Iy$, $U(y)=Uy$, $V(y)=Vy$
where $C_S, C_I, U, V>0$. Then the payoffs to the Invader and the
Scanner, if they use strategies $x\in [a,b]$ and $y\in [a,c]$
respectively, become:

\small
\begin{equation}
\label{eE_a}
\begin{split}
&v_I(x,y)=U(1-x-y)y-F(x+y)-C_Iy,\\
&\mbox{for known Invader's type:}\\
&v_S(x,y)= F(x+y)-Vy(1-x-y)-C_Sx, \\
&\mbox{for unknown Invader's type:}\\
&v^E_S(x,y)= q(F(x+y)-Vy(1-x-y))\\
&+(1-q)(F(x+a)-Vy(1-x-a))-C_Sx.
\end{split}
\end{equation}
\normalsize \noindent Note that linearized payoffs have found
extensive usage for a wide array of problems in wireless networks
\cite{AAG2011, SE2009, KRZ1999, KALKZ1999, GT2013}. Of course,
such approach simplifies the original problem  and only gives  an
approximated solution. Meanwhile it can also be very useful:
sometimes it allows one to obtain solution explicitly, and allows
one to look inside of the structure of the solution as well as the
correlation between parameters of the system.

\subsection{\label{sec:Best Response} Best response strategies for the Invader's reward related to bandwidth used}

In this section we give best response strategies for the Scanner
and the Invader, i.e. such strategies that
$\mbox{BR}_S(y)=\arg\max_x v_S(x,y)$ and
$\mbox{BR}_I(x)=\arg\max_y v_I(x,y)$.

\begin{Theorem}
\label{th_BR} In the second step of the game the Scanner and the
Invader have the best response strategies $\mbox{BR}_S(y)$ and
$\mbox{BR}_I(x)$ are given as follows:

\small
\begin{equation}
\label{e_th_BR_1}
 \mbox{BR}_S(y)=
\begin{cases}
a,& y<(C_S-F)/V,\\
\mbox{any from } [a,b],&  y=(C_S-F)/V,\\
b,&  y>(C_S-F)/V,
\end{cases}
\end{equation}

\begin{equation}
\label{e_th_BR_2}
\begin{split} \mbox{BR}_I(x)=
\begin{cases}
a,& T-2a\leq x,\\
L(x),&T-2c<x<T-2a,\\
c,&  x\leq T-2c
\end{cases}
\end{split}
\end{equation}
\normalsize \noindent
with
\small
\begin{equation}
\label{e_th_BR_2_L} L(x)=(T-x)/2, \quad T= (U-F-C_I)/U.
\end{equation}
\normalsize
\end{Theorem}

\subsection{\label{sec:Nash Equilibrium}  Equilibrium: the Invader's Award  is Related to the Width of
Band in Use}

\begin{figure}
\centering
\includegraphics[width=0.24\textwidth]{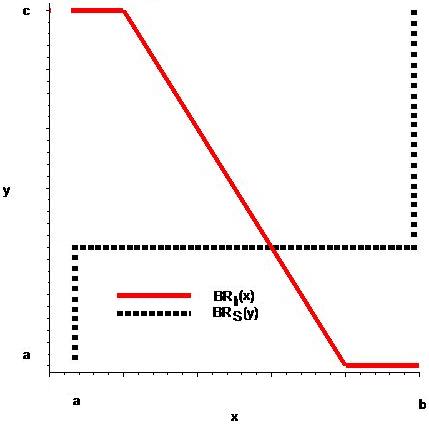}
\caption{\label{Fig_0} The Nash equilibrium as an intersection of
the best response curves}
\end{figure}

The equilibrium  in the second step of the problem exists since
the payoff to the Scanner is linear on $x$ and the payoff to the
Invader is concave on $y$.    The equilibrium can be found by
(\ref{e_NE}) as a couple of strategies $(x,y)$ which are the best
response to each other, i.e. $x=\mbox{BR}_S(y)$ and
$y=\mbox{BR}_I(x)$, i.e. the intersection of the best response
curves (Figure~\ref{Fig_0}). Such intersection always exists and is
unique as shown in the following theorem.

\begin{Theorem}
\label{th_2} Let the Invader's award  be related to the width of
band in use.  This game has unique Nash equilibrium  in the second
step, and it is given by Table~\ref{Tbl1} with $R=(C_S-F)/V.$
\small
\begin{center}
\begin{table*}[ht]
{\small \hfill{}
\begin{tabular}{|c|c|c|c|c|c|}
\hline\hline
\textbf{Case}&\textbf{Condition}&\textbf{Condition}&$x$&$y$&\textbf{Detection
probability}\\\hline\hline
$i_1$&$R<a$&$L(b)<a$&$b$&$a$&$a+b$\\\hline $i_2$&$R<a$&$a\leq
L(b)\leq c$&$b$&$L(b)$&$b+L(b)$\\\hline $i_3$&$R<a$&$c<
L(b)$&$b$&$c$&$c+b$\\\hline
$i_4$&$R>c$&$L(a)<a$&$a$&$a$&$2a$\\\hline $i_5$&$R>c$&$a\leq
L(a)\leq c$&$a$&$L(a)$&$a+L(a)$\\\hline $i_6$&$R>c$&$c<
L(a)$&$a$&$c$&$a+c$\\\hline $i_7$&$a\leq R\leq c$&$L(b)\leq R\leq
L(a)$&$L^{-1}\left(R\right)$&$R$&$L^{-1}\left(R\right)+R$\\\hline
$i_8$&$a\leq R\leq c$&$L(a)\leq a$&$a$&$a$&$2a$\\\hline
$i_9$&$a\leq R\leq c$&$a<L(a)<R$&$a$&$L(a)$&$a+L(a)$\\\hline
$i_{10}$&$a\leq R\leq c$&$c<L(b)$&$b$&$c$&$c+b$\\\hline
$i_{11}$&$a\leq R\leq
c$&$R<L(b)<c$&$b$&$L(b)$&$b+L(b)$\\\hline\hline
\end{tabular}
} \hfill{}  \caption{\label{Tbl1} The Nash equilibrium $(x,y)$
with $L^{-1}(R)=L^{-1}\left((C_S-F)/V\right)=T-2(C_S-F)/V$.}
\end{table*}
\end{center}
\normalsize
\end{Theorem}

\subsection{\label{sec:Nash Equilibrium_Unknown}
Equilibrium: Unknown whether  the Invader's reward  is related to bandwidth used}

For marginal  probabilities of the problem with unknown
Invader's type (where indeed there is complete confidence in the
Invader's type) we already have the solution. Namely, if $q=1$ then
the Invaders' reward depends on width in use. So, the equilibrium
is unique and given by Theorem~\ref{th_2}. If $q=0$ then the
Invaders' reward does not depend on width in use. Thus, the
equilibrium is unique again and it is $(\mbox{BR}_S(a),a)$. For
the inside probabilities the equilibrium is given by the following
theorem

\begin{Theorem} \label{th_3}  Let it be unknown whether the Invader's award  is related to the bandwidth used.  This game has unique Nash equilibrium  in the second
step, and it is given by Table~\ref{Tbl1} with
$R=(C_S-F-(1-q)Va)/(qV)$ and $L^{-1}(R) =T-2(C_S-F-(1-q)Va)/(qV)$.
\end{Theorem}

\section{\label{sec:Numerical illustrations}Numerical illustrations}

\begin{figure}
\centering
\includegraphics[width=0.24\textwidth]{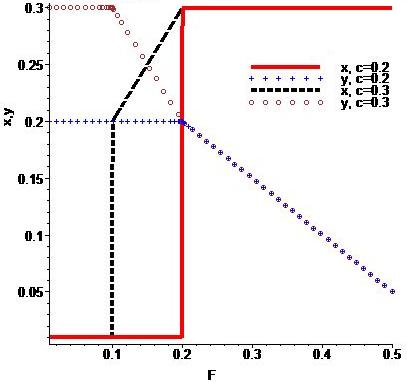}
\caption{\label{Fig_1} Equilibrium strategy $x$ and $y$ when the
Invader's reward depends on the bandwidth in use}
\end{figure}

\begin{figure}
\centering
\includegraphics[width=0.24\textwidth]{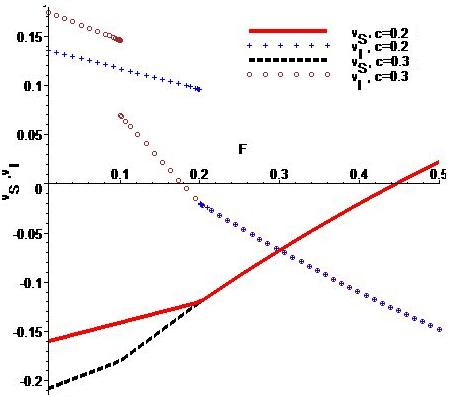}
\caption{\label{Fig_2} The equilibrium payoffs  $v_S$ and $v_I$
when the Invader's reward depends on the bandwidth in use}
\end{figure}

\begin{figure}
\centering
\includegraphics[width=0.24\textwidth]{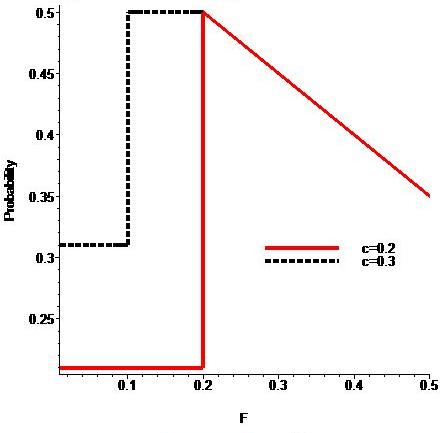}
\caption{\label{Fig_3} Probability of detection when the Invader's
reward depends on the bandwidth in use}
\end{figure}

As a numerical illustration of the scenario when the Invader's
reward depends on the bandwidth in use we consider $U=V=1$,
$a=0.01$, $b=0.3$,  $C_S=0.4$, $C_I=0.1$. Figures~\ref{Fig_1}
and~\ref{Fig_2} demonstrate the equilibrium strategies as a
function of the fine $F$ for the Invader's characteristics $c=0.2,
0.3$.
 Increasing fine makes the Scanner and the Invader employ larger
 and smaller bands correspondingly. The Scanner would alter his strategy by a sudden jump
 at a threshold value ($F=0.1(0.2)$ for $c=0.2(0.3)$). In spite of the fact that
 the Invader would vary his strategy continuously, his payoff experiences a sudden drop.
 This also leads to an increase in detection probability by a jump. Since
 the Scanner already gets the upper band, while the Invader still does not
 get to the lower band, further increasing of the fine leads to
 continuous decreasing of the detection probability due to
 the smaller bandwidth employed by the Invader (Figure~\ref{Fig_3}).

\begin{figure}
\centering
\includegraphics[width=0.24\textwidth]{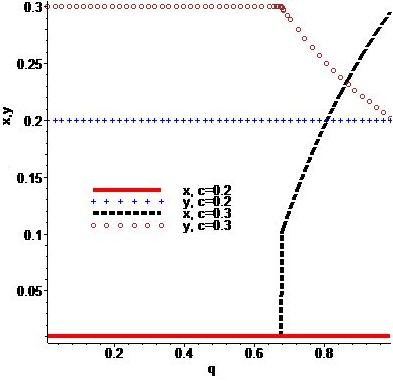}
\caption{\label{Fig_4} Equilibrium strategy $x$ and $y$ with
uncertainty about the Invader's award}
\end{figure}

\begin{figure}
\centering
\includegraphics[width=0.24\textwidth]{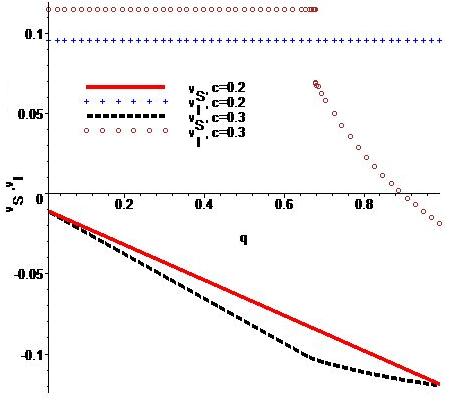}
\caption{\label{Fig_5} The equilibrium payoffs  $v_S$ and $v_I$
with uncertainty about the Invader's award}
\end{figure}

Figure~\ref{Fig_4} and~\ref{Fig_5} demonstrate the equilibrium
strategies as a function of probability $q$ (i.e. uncertainty
about the Invader's reward type) and fine $F=0.2$. The result
essentially depends on the Invader's characteristics. For  $c=0.2$
the equilibrium strategies do not depend on the probability $q$.
For $c=0.3$ they are constant for $q<0.68$, while increasing of
the Scanner strategy by a jump at switching point $q=0.68$ drops
also the Invader's payoff by a jump. This ``jumping nature'' for
the Invader's payoff means that when assigning a fine one has to
take into account the fact that sometimes a simple increase in
fine might not change the Invader's behavior, but that once a
threshold value is hit, his behavior will change dramatically.

\begin{figure}
\centering
\includegraphics[width=0.24\textwidth]{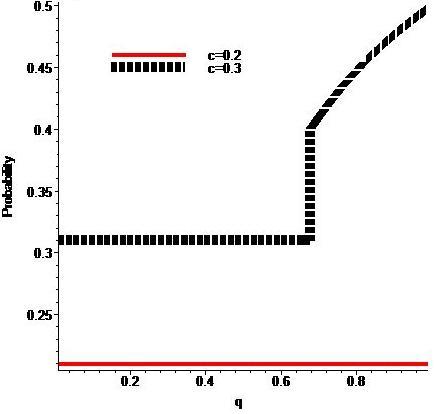}
\caption{\label{Fig_6} Probability of detection  with uncertainty
about the Invader's award}
\end{figure}

\section{\label{sec:Discussion}Discussion}

In this paper we suggest a simple model of finding the optimal
bandwidth to scan for detection of an Invader and found the
solution of this model explicitly.  We have shown that the optimal
width essentially depends on the scenario and under some
conditions a small varying of network parameters and fine could
lead to jump changes in the optimal strategies, as well as in the
payoffs of the rivals. This mixture between continuous and
discontinuous behavior of the Invader under the influence of fine
implies that the network provider has to carefully make a value
judgement: some threshold values of fine could have a huge impact
on the Invader, while in the other situations a small increase
will have a minimal impact on the strategies used. A goal for our
future investigation, of course, is to investigate the
non-linearized detection probability. Also, since our
investigation showed that the optimal scanning essentially depends
on the Invader's characteristics, we intend to extend our model to
the case of incomplete information on the characteristics as well
as to multi-scanner \cite{Comaniciu2005, G1997} or multi-invader
systems \cite{AAG2009a}.

\section{\label{sec:Appendix}Appendix}

\subsection{Proof of Theorem~\ref{th_1}}

Suppose that the Invader uses a band $B_I$ with width $y$ and the
Scanner with equal probability employ a band  from the set
$A_{-S}$ ($A_{+S}$) for $1-(x+y)M\leq y$ (for $1-(x+y)M>y$). The
intervals composing $A_{-S}$ and $A_{+S}$ are separated from each
other by at most $y$. So, at least one band from $A_{-S}$ for
$1-(x+y)M\leq y$ and from $A_{+S}$ for $1-(x+y)M> y$ intersects
with $B_I$.  So, detection probability is greater or equal to
$1/M$ for $1-(x+y)M\leq y$ and it is is greater or equal to
$1/(M+1)$ for $1-(x+y)M>y$.

Suppose that the Scanner uses a band $B_S$ with width $x$ and the
Invader with equal probability employ a band  from the set
$A_{-I}$ ($A_{+I}$) for $1-(x+y)M\leq y$ (for $1-(x+y)M>y$). The
intervals composing $A_{-I}$ and $A_{+I}$ are separated from each
other by more that $x$. So, at most one band from $A_{-I}$ for
$1-(x+y)M\leq y$ and from $A_{+I}$ for $1-(x+y)M> y$ intersects
with $B_S$.  So, detection probability is less or equal to $1/M$
for $1-(x+y)M\leq y$ and it is is less or equal to $1/(M+1)$ for
$1-(x+y)M>y$ and the result follows.
 \cqfd

\subsection{Proof of Theorem~\ref{th_BR}} Note that $v_S(x,y)=x(F+Vy-C_S)+y(F-V+Vy)$. So, for a fixed
$y$ the payoff $v_S(x,y)$ is linear on $x$. Thus,
$\mbox{BR}_S(y)=\arg\max_x v_S(x,y)$ is defined by sign of
$F+Vy-C_S$ as it is given by (\ref{e_th_BR_1}).

Note that $v_I(x,y)=(U(1-x)-F-C_I)y^2-Uy^2-xF$. So, for a fixed
$x$ the payoff $v_I(x,y)$ is a concave quadratic polynomial on $y$
getting its absolute maximum at $y=(U(1-x)-F-C_I)/(2U)$. Thus, the
maximum of $v_I(x,y)$ within $[a,c]$ is reached either on its
bounds $a$ and $c$ or at $y=(U(1-x)-F-C_I)/(2U)$ if it belongs to
$[a,c]$ as it is given by (\ref{e_th_BR_2}). \cqfd

\subsection{Proof of Theorem~\ref{th_2}} First note that $(x,y)$ is a Nash equilibrium
if and only if it is a solution of equations $x=\mbox{BR}_S(y)$
and $y=\mbox{BR}_I(x)$ with $\mbox{BR}_S(y)$ and $\mbox{BR}_I(x)$
given by Theorem~\ref{th_BR}.

By (\ref{e_th_BR_2_L}) we have that (\ref{e_th_BR_2}) is
equivalent to

\begin{equation}
\label{e_Equiv}
\begin{split} \mbox{BR}_I(x)=
\begin{cases}
a,& L(x)\leq a,\\
L(x),&a<L(x)<c,\\
c,&  c\leq L(x).
\end{cases}
\end{split}
\end{equation}

Let $a>(C_S-F)/V$. By (\ref{e_th_BR_1}) $\mbox{BR}_S(y)\equiv b$.
This, jointly with (\ref{e_Equiv}),  implies ($i_1$)-($i_3$).

Let $(C_S-F)/V>c$. By (\ref{e_th_BR_1}) $\mbox{BR}_S(y)\equiv a$.
Then (\ref{e_Equiv})  implies ($i_4$)-($i_6$).

Let $a\leq (C_S-F)/V\leq c$. First note $L(x)$ is linear
decreasing function from $L(a)$ for $x=a$ to $L(b)$ for $x=b$.

\begin{description}
\item[(a)] Let $L(b)\leq (C_S-F)/V\leq L(a)$. Then the equation
$L(x)=(C_S-F)/V$ has the unique root within $[a,b]$. Thus,
(\ref{e_th_BR_1}) and (\ref{e_Equiv}) yield ($i_7$).

\item[(b)] Let $L(a)\leq (C_S-F)/V$. Then, $L(x)<(C_S-F)/V$ for $x\in (a,b]$.
So, by (\ref{e_th_BR_2}), $\mbox{BR}_I(x)<c$ for $x\in [a,b]$.
Besides, by the assumption, the equation $L(x)=(C_S-F)/V$ does not
has root in $[a,b]$. Thus, by (\ref{e_th_BR_1})
$\mbox{BR}_S(y)\equiv a$. So, (\ref{e_Equiv})  implies ($i_8$) and
($i_9$).

\item[(c)] Let $(C_S-F)/V<L(b)$. Then $L(x)>(C_S-F)/V$ for $x\in [a,b)$.
Thus, by (\ref{e_th_BR_2}), $\mbox{BR}_I(x)>a$ for $x\in [a,b]$.
Besides, by the assumption, the equation $L(x)=(C_S-F)/V$ does not
has root in $[a,b]$. So, by (\ref{e_th_BR_1}), $\mbox{BR}_S(y)=b$,
and, (\ref{e_Equiv})  implies ($i_{10}$) and ($i_{11}$).
\end{description}

\cqfd

\subsection{Proof of Theorem~\ref{th_3}}
Note that
$v^E_S(x,y)=(F-C_S+(1-q)Va+Vqy)x+qy(F-V+Vy)+(1-q)a(F-V+Va)$. So,
for a fixed $y$ the payoff $v^E_S(x,y)$ is linear on $x$. Thus,
$\mbox{BR}_S^E(y)=\arg\max_x v_S^E(x,y)$ is given as follows

\begin{equation*}
\mbox{BR}_S^E(y)
 =\begin{cases}
a,& y<R,\\
\mbox{any from } [a,b],&  y=R,\\
b,&  y>R
\end{cases}
\end{equation*}
with $R=(C_S-F-(1-q)Va)/(qV)$. This, jointly with
Theorem~\ref{th_BR} and the proof of Theorem~\ref{th_2}, implies
the result. \cqfd

\end{document}